\documentclass[aps,prc,twocolumn,showpacs,preprintnumbers,amsmath,amssymb,superscriptaddress]{revtex4-1}

\usepackage{graphicx}
\usepackage{epsfig}
\usepackage{dcolumn}
\usepackage{bm}
\usepackage{color}

\begin{document}


\title{ Global trends of nuclear $d_{5/2}^{2,3,4}$ configurations: Application of a simple effective-interaction model}
\author{M.~Wiedeking}
\email{wiedeking@tlabs.ac.za}
\affiliation{iThemba LABS, P.O. Box 722, 7129 Somerset West, South Africa}
\affiliation{School of Physics, University of the Witwatersrand, Johannesburg 2050, South Africa}
\author{A.~O. Macchiavelli}
\email{aomacchiavelli@lbl.gov}
\affiliation{Lawrence Berkeley National Laboratory, Berkeley, California 94720, USA}


\date{\today}

\begin{abstract}

With new experimental information on nuclei far from stability being available, a systematic investigation of excitation energies and electromagnetic properties along the $N=10, 11, 12$ isotones and $Z=10, 11, 12$ isotopes is presented. The experimental data are discussed in the context of the appearance and disappearance of shell closures at $N=Z=8,14,16,20$, and compared to an effective-interaction approach applied to neutrons and protons 
in $d_{5/2}^{2,3,4}$ configurations. In spite of its simplicity the model is able to explain the observed properties.

\end{abstract}

\maketitle

\section{Introduction}

The interaction between valence neutrons and protons plays a prominent role in understanding how nuclear structure evolves with changing nucleon numbers~\cite{Hamamoto65,Federman79,Federman84,Casten85, Casten96}. The monopole average of the neutron-proton interaction can cause large changes in the effective single-particle energies (ESPE) that may significantly alter the underlying shell structure,  allowing the pairing and quadrupole components to develop superfluidity and deformation  away from closed shells~\cite{Poves87,Warburton90,Otsuka01,Zuker00}. Indeed, changes in the ESPE's can modify the available valence space and thus influence the onset and strength of collective modes (e.g. quadrupole versus pairing). Measuring and understanding the evolution of ESPE's from stability to the drip lines is essential for a complete understanding of nuclear structure effects and has been a central theme of study for the last two decades~\cite{Otsuka20}. With rapid advances in detector systems, exotic beams production, and large-scale computing it is now possible to  investigate the evolution of shell-structure and collectivity for long sequences of nuclei in isotopic or isotonic chains reaching far away from stability.
Trends obtained from such sequences often contradict our understanding from stable nuclei, implying that the observed differences are due to the isospin dependence of the nuclear force, notably the tensor component~\cite{Otsuka01}.
Studies on the emergence (erosion) of new (traditional) shell gaps have featured predominantly in experimental and theoretical nuclear structure physics research in nuclei away from the line of $\beta$ stability. 

In light nuclei, examples of changes in the shell structure include but are not limited to i) the emergence of new shell closures \cite{Jan2005, Sor2008} at $N=Z=14$ \cite{Thi2000, Cot2007} and $N=Z=16$ \cite{Dlo2003, Hof2009}, ii) the weakening of the proton $p-sd$ shell closure as neutrons are added to $Z=8$ isotopes \cite{Wie2005}, and iii) the island of inversion at $N=20$, where $2p2h$, $4p4h$, ...  excitations from the $sd$-shell to intruder orbits from the $fp$ shell become energetically accessible and yield deformed ground-state structures \cite{Thi1975, Det1979, Poves87,Warburton90}.

Fundamental quantities in even-even nuclei such as the excitation energy of the first-excited $2^+$ state $E(2^+_1)$, the electric quadrupole transition strength $B(E2,2^+_1 \rightarrow 0^+ )$, and the ratio $R_{42}=E(4^+_1)$/$E(2^+_1)$ with $E(4^+_1)$ being the energy of the first-excited $4^+$ state, have been shown to provide important insight into the structural evolution of nuclei~\cite{Mall1959, Casten16}.

For instance, the scale of $E(2^+_1)$ and $R_{42}$ already provide information on the collectivity and allow for a classification of nuclear shapes. An increase in $B(E2,2^+_1 \rightarrow 0^+ )$, compared to that of a single-proton excitation, suggests that more active nucleons contribute coherently to the transition and signals deformations (dynamic or static) of the nuclear shape~\cite{BM}.  
Similarly, the excitation energies and ordering of low-lying states in odd-A nuclei can also be effectively used to determine the migration and inversion of orbits as a function of the number of protons and neutrons. 

With the wealth of available experimental data in the $13<A<41$ region it is instructive to investigate trends of these fundamental quantities and correlate them to the number of available valence nucleons (holes) \cite{Casten96, Bontasos13} and to simple effective interactions and coupling configurations. The latter approach has been  proposed and discussed in detail by Talmi \cite{Talmi62, Talmi93} and is applied here to describe general properties of $d_{5/2}^n$ configurations. 

In section II we compile the available experimental data on $E(2^+_1)$, $E(4^+_1)$, and $B(E2,2^+_1 \rightarrow 0^+ )$, as well as the energies of the first-excited  3/2$^+$ ($E(3/2^+_1$)),  5/2$^+$ ($E(5/2^+_1$)), and  9/2$^+$ ($E(9/2^+_1$)) states for $Z=N=10,11,12$ isotopes and isotones. The effective interaction model is briefly reviewed in section III. A comparison with the data from $d_{5/2}^{2,3,4}$ configurations, the discussion of the results and the connection to the $N_pN_n$ scheme~\cite{Casten96} are presented in section IV.

\section{Experimental Data}

\begin{figure}
\includegraphics[clip,angle=0,width=\columnwidth]{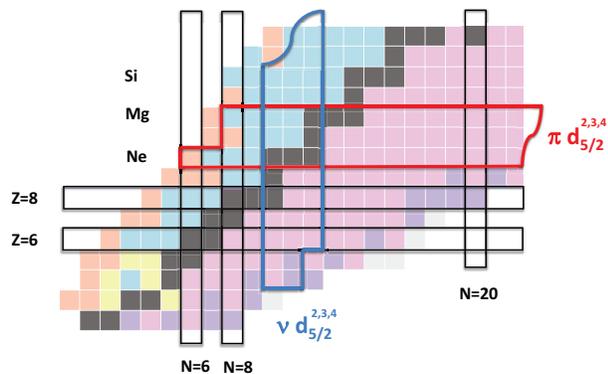}
\vspace{-0.8cm}
\caption{(Color online) Region of the Segr\`e Chart showing the isotones (blue) and isotopes (red) considered in the $d_{5/2}^{2,3,4}$ analysis.}
\label{fig:first}
\end{figure}

Experimental data are available for a large number of nuclei in $d_{5/2}^{2,3,4}$ configurations.  Fig.~\ref{fig:first} shows the isotopes and isotones which have been considered in this work.

Values of $E(2^+_1)$, $E(4^+_1)$, and $B(E2,2^+_1 \rightarrow 0^+ )$ are summarized in Table \ref{tab:10-12} for  the $N=10,12$ and $Z=10,12$ even-even isotones and isotopes, while Table \ref{tab:11} gives the experimental values of $E(3/2^+_1)$, $E(5/2^+_1)$, and E$(9/2^+_1)$ in the $Z=N=11$ odd-A isotopes and isotones.  This information is also presented in graphical form in Figs.~\ref{fig:eveneven} and \ref{fig:evenodd}.

\begin{table*}
\begin{center}
\caption{\label{tab:10-12}Energies of the first 2$^+$ ($E(2^+_1)$) and 4$^+$ ($E(4^+_1$)) states, the ratio $R_{42}=E(4^+_1)/E(2^+_1)$, the half-life T$_{1/2}$ (when the $B(E2,2^+_1 \rightarrow 0^+ )$ is calculated from it), and the electric quadrupole transition strength $B(E2,2^+_1 \rightarrow 0^+ )$ for neutrons and protons in the $d_{5/2}^2$ and $d_{5/2}^4$ configurations. The data is compiled from evaluations as indicated by the references in the first column except for instances where data is taken from other sources as included next to the relevant entries. Blank entries are indicative of the absence of experimental data.}
\begin{tabular}{c c  c  c  c  c   c}
\hline
 			&Nucleus					& $E(2^+_1)$ (keV)			& $E(4^+_1)$ (keV)			& $R_{42}$				& T$_{1/2}$ (ps)				&$B(E2,2^+_1 \rightarrow 0^+ )$ (W.U.)\\
\hline
			&$^{14}$Be \cite{Ajz1991}		& 1540(13) \cite{Sug2007}		& 						&  								& 							&	\\
			&$^{16}$C \cite{Kel1993}		& 1760(2)$^a$ 				& 4129.5(45)$^a$ 				&  2.3463(37)						& 8.0($^{+15}_{-18}$)$^a$				& 1.75($^{+33}_{-39}$)	\\
			&$^{18}$O \cite{Til1995}		& 1982.07(9)				& 	3554.84(40)			&  1.79350(22)						& 1.94(5)						& 3.40(9)	\\
N=10 isotones	&$^{20}$Ne \cite{Til1998}		& 1633.674(15)				& 	4247.7(11)				&  2.60009(67)						& 		0.73(4)				& 20.7(1.1)\\
			&$^{22}$Mg \cite{Bas2015}	& 1247.02(3)				& 	3308.22(6)				&  	2.652901(80)						& 						& 20.8(26)$^\dagger$ \cite{Hend2018}\\
			&$^{24}$Si \cite{Fir2007}		& 1874(3) \cite{Wolf2019}		& 3471(6)$^b$ \cite{Wolf2019}		&  	1.8522(44)					& 							& 4.7(15)$^\dagger$ \cite{Kan2002}		\\
\hline
			&$^{18}$C \cite{Kelley2017}		& 1585(15)$^c$				& 4000(32)$^\star$				& 2.524(31) 						& 14.3(28))$^c$				&1.41(28)\\
			&$^{20}$O \cite{Til1998}		& 1673.68(15)				& 3570.5(9)$^d$  			& 2.13332(57) 						& 7.3(3)						& 1.83(10)	\\
			&$^{22}$Ne \cite{Bas2015}		& 1274.537(7)				& 3357.2(5) 				& 2.63406(39) 						& 3.60(5) 						& 12.76(18)	\\	
N=12 isotones	&$^{24}$Mg \cite{Fir2007}		& 1368.672(5)				& 4122.889(12)				& 3.012328(14)						& 1.33(6)						& 21.5(10)	\\	
			&$^{26}$Si \cite{BasHur2016}	& 1797.3(1)				& 4446.37(18)$^{be}$			& 2.47392(17) 						& 0.440(40)					& 15.0(14)\\
			&$^{28}$S \cite{Bas2013}		& 1507(7)					& 						&  								& 							& 7.2(12)$^\dagger$ \cite{Tog2012}\\
			&$^{30}$Ar \cite{Singh2016}	&  530(22)$^b$ \cite{Xu2018}	&						&								&							&\\
\hline
			&$^{16}$Ne \cite{Kel1993}		& 1733(53)$^f$ 			& 						&  								& 							&	\\
			&$^{18}$Ne \cite{Til1995}		& 1887.3(2)				& 3376.2(4)				&  1.78890(28)						& 0.46(4)					& 18.3(16)	\\
			&$^{20}$Ne \cite{Til1998}		& 1633.674(15)				& 	4247.7(11) 			&  2.60009(67)						& 		0.73(4)			& 20.7(1.1)\\
			&$^{22}$Ne \cite{Bas2015}		& 1274.537(7)				& 3357.2(5) 				& 2.63405(39) 						& 3.60(5) 					& 12.76(18)	\\
Z=10 isotopes	&$^{24}$Ne \cite{Fir2007}		& 1981.6(4)				& 3962.0(8) \cite{Hof2003}		& 1.99939(57) 						& 0.66(15)					& 6.8(16)	\\
			&$^{26}$Ne \cite{BasHur2016}	& 2018.28(10)				& 3516.8(5) \cite{Lep2013}			& 1.74247(26) 						& 0.60(8)					& 6.16(82)	\\
			&$^{28}$Ne \cite{Bas2013}		& 1304(3)					& 	3010(6)$^b$			& 2.3083(70)						& 				& 5.23(91)$^\dagger$ \cite{Iwa2005}\\
			&$^{30}$Ne \cite{Bas2010}		& 792(4)					& 2235(12)$^b$				& 2.822(21)							& 				& 10.0(29)$^\dagger$ \cite{Doo2016}\\
			&$^{32}$Ne \cite{Oue2011}	& 715.5(106)$^{bg}$			& 	2119(19)$^b$ \cite{Mur2019}					&  	2.962(51)							& 						&	\\
\hline
			&$^{20}$Mg \cite{Kell2019}	& 1598(10)				& 	3700(20)$^b$				&  	2.315(19)							& 						&	\\
			&$^{22}$Mg \cite{Bas2015}	& 1247.02(3)				& 	3308.22(6)				&  	2.652901(80)						& 					& 20.8(26)$^\dagger$ \cite{Hend2018}	\\
			&$^{24}$Mg \cite{Fir2007}		& 1368.672(5)				& 4122.889(12)				& 3.012328(14)						& 1.33(6)					& 21.5(10)	\\		
			&$^{26}$Mg \cite{BasHur2016}	& 1808.74(4)				& 4318.89(5)				& 2.387789(60)						& 0.476(21)				& 13.42(59)	\\	
			&$^{28}$Mg \cite{Bas2013}	& 1473.54(10)				& 4021.0(5)				& 2.72880(39) 						& 1.07(13)$^h$				& 15.1(19) \\
Z=12 isotopes	&$^{30}$Mg \cite{Bas2010}	& 1482.8(3)				& 3379.0(8) \cite{Dea2010}		& 2.27880(71) 						& 1.5(2)					& 9.5(13)	\\
			&$^{32}$Mg \cite{Oue2011}	& 885.3(1)					& 2322.3(3) 				& 2.62318(45) 						& 11.4(20)					& 14.7(16)$^i$\\
			&$^{34}$Mg \cite{Nic2012}		& 660(7)					& 2047(16)$^b$ \cite{Doo2013}	& 3.102(41) 						& 40(8)					& 17.3(35)\\
			&$^{36}$Mg \cite{Nic2012_2}    	& 660.6(78)$^j$				& 2014(32)$^k$ 			& 3.049(60) 						& 						&  15.0(34)$^\dagger$ \cite{Doo2016}	\\
			&$^{38}$Mg \cite{Chen2018}	 & 645.5(60)$^l$			        & 1991.5(205)$^l$   & 3.085(43) 			& 			&	\\
			& $^{40}$Mg \cite{Chen2017} & 500(14)$^{b}$ \cite{Cra2019}   	& 			        & 						& 			&	\\
\hline
\end{tabular}

\label{tab:neutrons}
\end{center}
$^\dagger$ Converted to Weisskopf Units with $B(E2)_{sp}=5.94~10^{-6} A^{4/3} e^2b^2$.
\newline
$^a$ Average of values measured with high-purity Germanium detectors reported in \cite{Wie2008} and \cite{Pet2012}.
\newline
$^b$ Tentative spin and parity assignment.
\newline
$^c$ Average from values reported in \cite{Ong2008} and \cite{Vos2012}.
\newline
$^\star$ Spin and parity has not been assigned and it is only assumed that this is the $4^+$ state c.f. \cite{Sta2008}.
\newline
$^d$ Average from values reported in \cite{Wie2005} and \cite{Sum2006}.
\newline
$^e$ A tentative assignment of 3842.2 (18) for the 4$^+_1$ state is given in the evaluation \cite{BasHur2016} but its existence is doubtful since the level is not observed in fusion evaporation \cite{Sew2007}, ($^3$He,n) \cite{Kom2014,Doh2015}, (p,t) \cite{Mat2010,Chi2010}, or fragmentation reaction data \cite{Che2012}. Instead the levels at 4446.37(18) is considered to be the 4$^+_1$ state \cite{Sew2007,Kom2014,Doh2015,Mat2010,Chi2010,Che2012} which has been corroborated \cite{Perez2016} 
\newline 
$^f$ Average from values reported in \cite{Muk2008,Wam2014,Bro2014}.
\newline
$^g$ Average from values reported in \cite{Oue2011} and \cite{Mur2019}.
\newline
$^h$ Average from values reported in the evaluation \cite{Bas2013} and measurement \cite{Tog2012}.
\newline
$^i$ Average of evaluated value from \cite{Oue2011} and \cite{Li2015}.
\newline
$^j$ Average from values reported in \cite{Nic2012_2,Doo2013,Mic2014,Doo2016,Cra2019}.
\newline
$^k$ Average from values reported in \cite{Kob2014} and \cite{Cra2019}.	
\newline
$^l$ Average from values reported in \cite{Doo2013,Chen2018,Cra2019}.
\newline

\end{table*}

\begin{figure}
\includegraphics[clip,angle=0,width=\columnwidth]{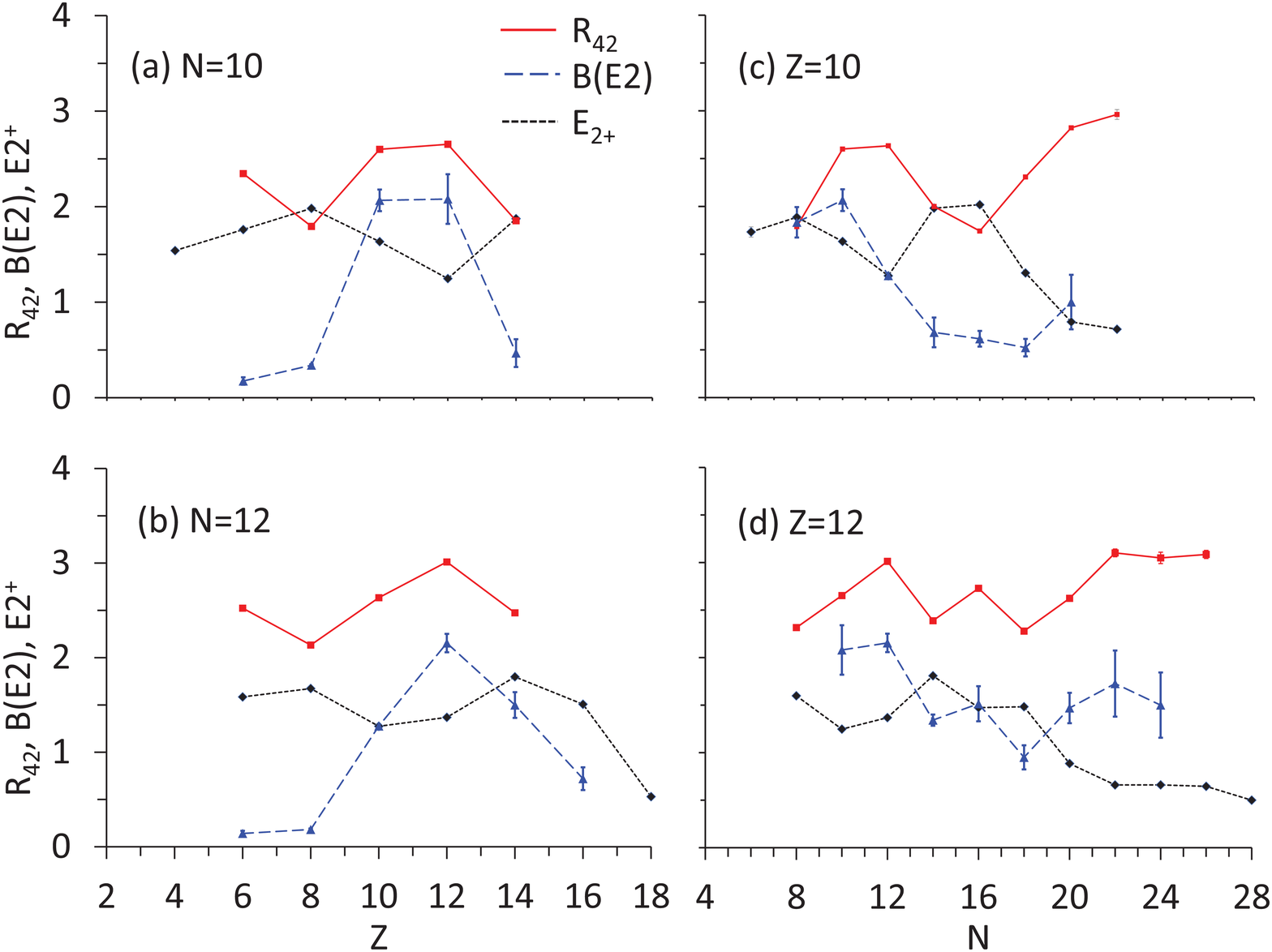}
\caption{\label{fig:eveneven} (Color online) Energies $E(2^+_1$) in MeV (black diamonds and dashed lines), ratios $R_{42}$ (red squares and solid lines), and $B(E2,2^+_1 \rightarrow 0^+ )$ in 10 W.U. (blue triangles and long-dashed lines) as a function of number of protons and neutrons in the $d_{5/2}^2$ and $d_{5/2}^4$ configurations for $N=10$ (a), $N=12$ (b), $Z=10$ (c), and $Z=12$ (d) nuclei.}
\label{fig:eveneven}
\end{figure}

\begin{table*}
\begin{center}
\caption{\label{tab:11} Energies of the first-excited 5/2$^+$ ($E(5/2^+_1)$), 3/2$^+$ ($E(3/2^+_1$)), and 9/2$^+$ ($E(9/2^+_1$)) states, the ratios $R_{39}=\frac{E(3/2^+_1)-E(5/2^+_1)}{E(9/2^+_1)-E(5/2^+_1)}$  and $R_{32}=\frac{E(3/2^+_1)-E(5/2^+_1)}{E(2^+_1)-E(0^+_1)}$ for neutrons and protons in the $d_{5/2}^3$ configuration. The $E(2^+_1)$ values in $R_{32}$ are averages of adjacent even-even nuclei where possible. Energies are compiled from published evaluations as indicated by the references in the first column. For instances where data is obtained from other sources, those are shown next to the relevant entries. Blank entries indicate the absence of data.}

\begin{tabular}{c c  c  c  c  c  c}
\hline
 			&Nucleus						& $E(5/2^+_1)$ (keV)		& $E(3/2^+_1)$ (keV) & $E(9/2^+_1)$ (keV) 		&  $R_{39}$ & $R_{32}$\\
\hline
			&$^{15}$Be \cite{Kel2015}				& 	0$^a$				& 1100$^{a,b}$ \cite{Kuc2015} & 			& 	& 		\\
			&$^{17}$C \cite{Kel2017}				& 332(2)				& 	0			&3085(25) &		 	-0.1206(13)	&-0.1985(22)
		\\
			&$^{19}$O \cite{Til1995}				& 	0				& 	96.0(5)		&2371.5(10)&		 0.04048(21)	&0.05252(27)
					\\
N=11 isotones	&$^{21}$Ne \cite{Fir2015}				& 	350.727(8)			& 		0	&2866.6(2)&			   -0.139406(12)	&-0.2411978(61)

	\\
			&$^{23}$Mg \cite{Fir2007_2}			& 	450.71(15)			& 	0	& 2713.3(7) \cite{Jenk2013} 				&  -0.199201(91)	&-0.34462(12)

				\\
			&$^{25}$Si \cite{Fir2009}				& 	0				& 	42.5(55)$^c$	&2365(7)$^a$ \cite{Long2018}		&  0.0180(23)	&0.02315(30)
			\\
			&$^{27}$S \cite{Bas2011}				& 		0$^a$			& 	$\leq$100$^a$ \cite{Gad2008}	& 			&	&0.066357(31)
		\\

\hline
			&$^{19}$Na \cite{Kel2015}				& 0$^a$				& 120(10)$^a$		&		& 0.0688(58)

			\\
			&$^{21}$Na \cite{Fir2015}				& 331.9(1)				& 0				&2829.1(7)		&-0.132909(55)	&	-0.230431(70)

			\\
			&$^{23}$Na \cite{Fir2007_2}			& 439.990(9)			& 0					&2703.500(25)	&  -0.1943839(46)	&-0.3329211(72)

			\\
			&$^{25}$Na \cite{Fir2009}				& 0					& 89.53(10)			&2418.5(8)$^a$ \cite{Von2015}	&  0.037019(43)	&0.047241(54)
			\\
Z=11 isotopes	&$^{27}$Na \cite{Bas2011}				& 0					& 62.9(6)$^a$ 	&2224.2(9)$^a$				&  0.02828(27)	&0.036027(34)

		\\
			&$^{29}$Na \cite{Bas2012}				& 72.0(5)$^{ad}$			& 0			&   					&&-0.05167(38)
	\\
			&$^{31}$Na \cite{Oue2013}			& 375(4)$^a$			& 0					&					&&-0.4471(52)
		\\
			&$^{33}$Na \cite{Che2011}			& 427(5)$^e$		& 0$^a$					&1875(19)$^a$ \cite{Doo2014}		&-0.2949(53)&	-0.620(14)
				\\
			&$^{35}$Na \cite{Che2011_2}			& 373(5)$^a$ \cite{Doo2014}	& 0$^a$			&						&&	-0.565(10)
	\\

\hline
\end{tabular}

\label{tab:protons}
\end{center}
$^a$ Tentative spin and parity assignment.
\newline
$^b$ The state and its energy has not been firmly accepted \cite{Kel2015}.
\newline
$^c$ Average from values reported in \cite{Rey2010} and \cite{Long2018}.
\newline
$^d$ Uncertainty not given in evaluated data base \cite{Bas2012} but discussed to be 0.5 keV or less \cite{Tri2006}.
\newline
$^e$ Average from values reported in \cite{Che2011} and \cite{Doo2014}.
\newline
\end{table*}

\begin{figure}
\includegraphics[clip,angle=0,width=\columnwidth]{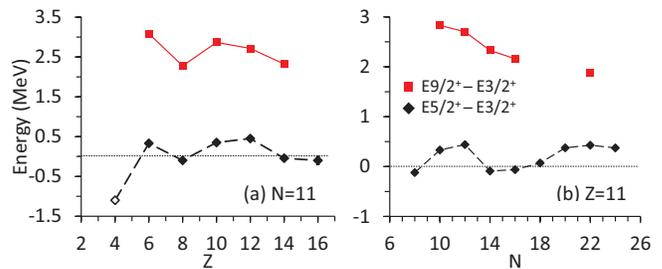}
\caption{\label{fig:evenodd} Energy differences $E(5/2^+_1)-E(3/2^+_1)$ (black diamonds)   and $E(9/2^+_1)-E(3/2^+_1)$ (red squares) for nucleons in the $d_{5/2}^3$ configuration for the $N=11$ (a) and $Z=11$ (b) nuclei.}
\label{fig:oddeven}
\end{figure}

\section{Effective Interaction Model}

Igal Talmi has discussed in depth the effective interaction model in $j^n$ configurations \cite{Talmi62, Talmi93}. The energy spacings and properties of low-lying states  are described by a two-body effective interaction obtained from experiment.
In spite of its simplicity the model seems to capture the main physical ingredients and is able to correlate well the empirical data. Here, we are specifically interested in the $(d_{5/2})^{2,3,4}$ configurations for which there are analytical solutions.  In these systems we can generate $I$= 0$^+$, 2$^+$, and 4$^{+}$ states for even-even 
and $J$=3/2$^+$, 5/2$^+$, and 9/2$^+$ states for odd-A nuclei.\\

\noindent
Following Talmi, given the 2-body matrix elements:
\begin{equation}
V_I=\langle{d_{5/2}^{2} I M} |V| d_{5/2}^{2} I M \rangle
\label{eq:matrixelement}
\end{equation}
\noindent the energies of the states in the odd-A nuclei can be readily obtained:
\begin{equation}
\begin{split}
E_{3/2}& = \frac{15}{7}V_2+\frac{6}{7}V_4  \\
E_{5/2}& = \frac{2}{3}V_0+\frac{5}{6}V_2+\frac{3}{2}V_4  \\
E_{9/2}& = \frac{9}{14}V_2+\frac{33}{14}V_4 
\end{split}
\label{eq:intenergies}
\end{equation}

\noindent
in terms of $V_I$ in Eq.~\ref{eq:matrixelement}.

We use a simple interaction consisting of short-range and long-range terms, written in the form:
\begin{equation}
V_I= -G\delta_{I,0} + \chi I(I+1)
\label{eq:pplusq}
\end{equation}

\noindent which could be considered as a simplified version of a Pairing-plus-Quadrupole force in a single-$j$ shell~\cite{Kumar}. In terms of the strengths, $G$ and $\chi$, we have for two particles (holes) in the $d_{5/2}$ orbit:
\begin{equation}
V_0 = -G \\  
V_2 = \frac{3}{5}\chi\\     
V_4 = 2\chi.
\label{eq:energies}
\end{equation}
\noindent
The results for the $d_{5/2}^3$ configurations follow from Eqs.~\ref{eq:intenergies} and \ref{eq:energies} and show that in a pairing dominated system ($\chi/G \ll$ 1) the paired 5/2$^+$ state is favored and the 3/2$^+$ and 9/2$^+$ unpaired states are degenerate. As the strength $\chi$ is increased relative to $G$, the 3/2$^+$ and 9/2$^+$ states move further apart, with the unpaired 3/2$^+$ state approaching the 5/2$^+$ state and becoming the ground stated for $\chi/G \gtrsim $ 1.4.  In an earlier work~\cite{Wie2008b} we applied this approach for the specific case of the transitional nucleus $^{18}$N and showed that $^{19}$O favors the $5/2^+$ paired neutron $d_{5/2}^3$ configuration for the  ground state in contrasts to $^{17}$C where the $3/2^+$ unpaired state is favored. \\

\begin{figure}
\includegraphics[clip,width=\columnwidth]{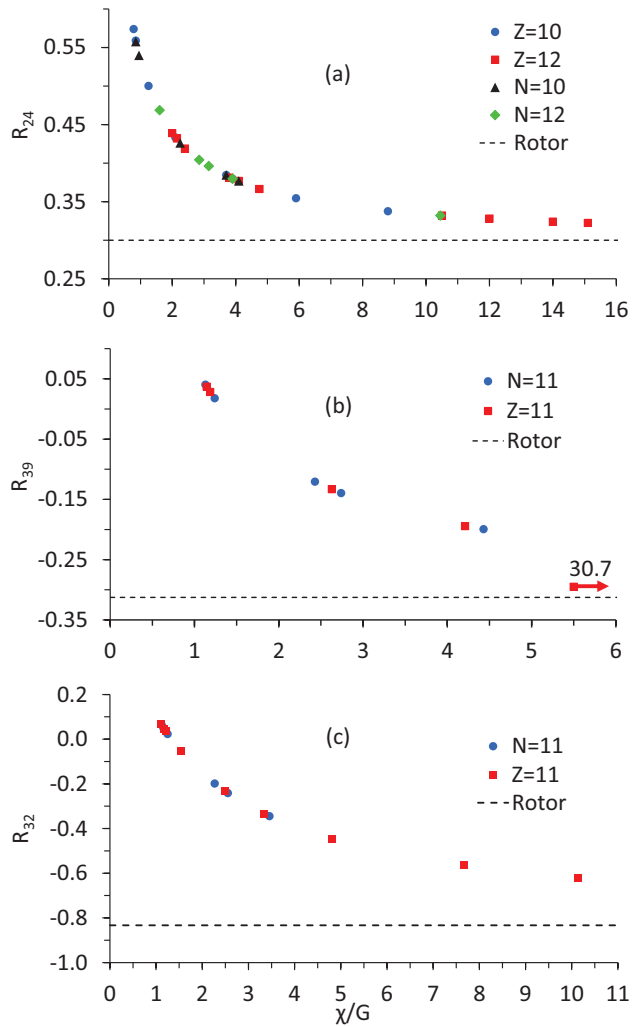}
\caption{(Color online) Ratios $R_{24}$, $R_{39}$, and $R_{32}$ versus $\chi/G$ for the $d_{5/2}^{2,4}$ configurations (a) and for the  $d_{5/2}^3$ configurations (b) and (c). For $^{33}$Na $\chi/G = 30.7$ in panel (b) and is indicated by a representative data point at $\chi/G =5.5$  with an arrow. The limits for a perfect rotor are indicated by horizontal dashed lines at $\frac{3}{10}$ for $R_{24}$, -$\frac{5}{16}$ for $R_{39}$, and -$\frac{5}{6}$ for $R_{32}$.}
\label{fig:model}
\end{figure}

\noindent 
Using the experimental data the $\chi/G$ values in the effective model can be extracted. To do so, it is convenient to present the data in terms of dimensionless energy ratios versus $\chi/G$: 
\begin{equation}
R_{24}=1/R_{42}=(E(2^+_1)-E(0_1^+)/(E(4^+_1)-E(0_1^+)) \nonumber
\end{equation}
for the even-even  $d_{5/2}^{2,4}$ configurations, 
\begin{equation}
R_{39}=(E(3/2^+_1)-E(5/2^+_1))/(E(9/2^+_1)-E(5/2^+_1)) \nonumber
\end{equation}
for the odd-A $d_{5/2}^3$ configurations, and 
\begin{equation}
R_{32}=(E(3/2^+_1)-E(5/2^+_1))/(E(2^+_1)-E(0^+_1)) \nonumber
\end{equation}
for a combination of odd-A $d_{5/2}^3$ and even-even $d_{5/2}^{2,4}$ configurations where the average $E(2^+_1)$ value from two adjacent even-even nuclei for a given odd-A nucleus was used where possible\footnote{The use of $R_{24}$, instead of the standard $R_{42}$, is to keep this in line with $R_{39}$ and $R_{32}$ since $R_{93}=1/R_{39}$ and  $R_{23}=1/R_{32}$ diverge at $\chi/G$=4/3.}.The results are shown in Fig.~\ref{fig:model}  for all isotones and isotopes under consideration. As a reference, the limits of the perfect rotor ($\chi/G \rightarrow \infty$) are also indicated at $\frac{3}{10}$ for $R_{24}$, -$\frac{5}{16}$ for $R_{39}$, and -$\frac{5}{6}$ for $R_{32}$.

When inspecting the data in Fig.~\ref{fig:model} it is apparent that the majority of nuclei with $d_{5/2}^{2,3,4}$ configurations reside within a transitional regime, stressing the fact that there is a delicate balance between the pairing and quadrupole forces in defining the structure of these nuclei.  The limits of a "perfect" rotor are approached primarily for $Z=11,12$ nuclei which lie in the $N=20$ island of inversion.   While the $d_{5/2}^{3}$ nucleus $^{33}$Na appears to exhibit the most established rotor behaviour, based on $R_{39}$, of any nucleus considered in this work, we have to keep in mind that the ($9/2^+$) spin assignment to the 1875(19) keV level is tentative and will be further discussed in the next section. 

\section{Results and Discussion}

Some interesting features are apparent when comparing chains of isotopes and isotones in the $d_{5/2}^{2,3,4}$ configurations. The nuclear properties under consideration in Fig. \ref{fig:eveneven} for the even-even nuclei
clearly show the $N=Z=8$ shell closure. Adding one or two pairs of protons or neutrons leads to an increase in collectivity before the $N=Z=14,16$ shell closures are reached. The $N=16$ shell closure at $Z=12$ shows $E(2_1^+), R_{42}$ and $B(E2,2^+_1 \rightarrow 0^+ )$ values which suggest a somewhat weakened $N=16$ gap compared to the $Z=10$ isotopes or $N=10,12$ isotones. 

For the lower-mass nuclei $^{16}$C and $^{18}$C the $B(E2,2^+_1 \rightarrow 0^+ )$ values appear suppressed which may imply magicity at $Z=6$ which is contradicted by the increase in $R_{42}$ and decrease in $E(2_1^+)$ values. This apparent conundrum has been discussed to be due to the states being dominated by neutron excitation and the small $B(E2,2^+_1 \rightarrow 0^+ )$ values are based on polarization effects \cite{Wie2008, Macc2014}.

Inspecting $^{28}$Ne and $^{30}$Mg a trend towards magicity is observed just prior to the rapid onset of deformation for $^{30}$Ne and $^{32}$Mg \cite{Iwa2005, Dea2010}. Indeed, the island of inversion at $N$=20 for the $Z=10,12$ isotopes is clearly visible with increasing $B(E2,2^+_1 \rightarrow 0^+ )$ and $R_{42}$ and decreasing $E(2_1^+)$ values  which remain relatively constant for $N>20$. 

It is important to note at this point that since we consider only simple configurations with the protons or neutrons occupying the $d_{5/2}$ orbit the effects of the "spectator" nucleons, due to the neutron-proton correlations, should be reflected in the behavior of the coupling constants.
Furthermore, given the short- and long-range components used in our schematic force in Eq. (3), it is appealing to correlate our empirically derived strengths with the number of valence particles ($N_p, N_n$) (holes) away from the shell closures, the so-called $N_p N_n$ scheme~\cite{Hamamoto65,Casten85,Casten96}.

\begin{figure}
\includegraphics[clip,angle=0,width=\columnwidth]{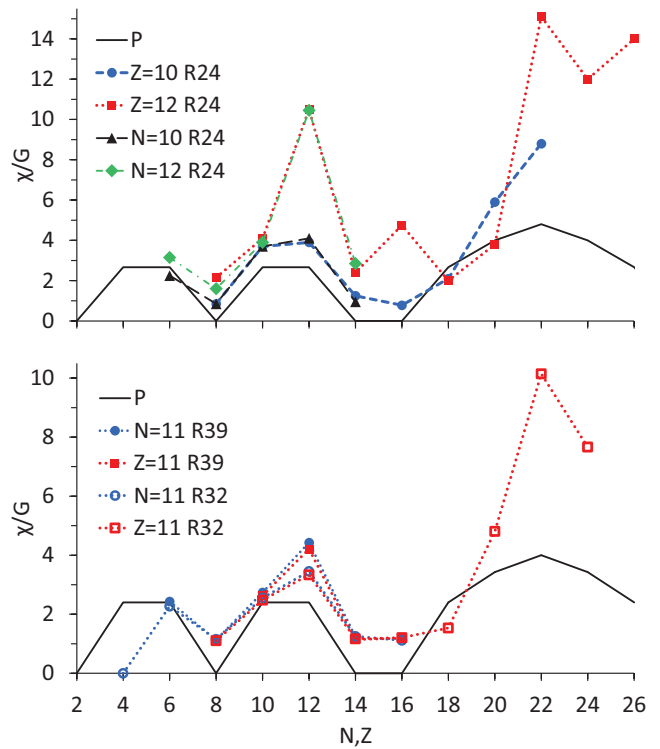}
\vspace{0cm}
\caption{(Color online) Trend of the $\chi/G$ values from $R_{24}$  of $d_{5/2}^{2,4}$ protons (blue solid circles and red solid squares) and neutrons (black solid triangles and green solid diamonds) (a) and from $R_{39}$ and $R_{32}$ from a combination of $d_{5/2}^{2,4}$ and $d_{5/2}^{3}$ protons (red solid and open squares) and neutrons (blue solid and open circles) (b). In each panel the scaled $N_p N_n$-scheme, $P$, (solid black line) with shell closures at $N=Z=2, 8, 14, 16$, and 28 is also displayed. Significant uncertainly exists for the data point at $Z=4$ in panel (b) and the data point $\chi/G$ = 30.7 for $^{33}$Na has been omitted (see text for details). The very good agreement observed in the $\chi/G$ values for some of the $N=Z=11$, $N=Z=10$, and $N=Z=12$ nuclei is naturally due to mirror symmetry.}
\label{fig:systematic}
\end{figure}

In fact, Fig.~\ref{fig:systematic} shows that the $\chi/G$ values consistently decrease for nuclei at the 8, 14, and 16 shell closures with the notable exception of $^{28}$Mg at $N=16$. Away from closed shells, where the long-range force gains importance, larger $\chi/G$ values 
are observed except for $^{15}$Be.  
 
For the $N=Z=11$ nuclei the proton-neutron valence interaction plays a key role in determining the excited and ground-state properties and the inversion from a paired to unpaired neutron (proton) coupling scheme due to proton-neutron interaction and can be quantitatively explained by the model of effective interactions. With three valence neutrons (protons) in the $j=5/2$ orbit the simplest neutron (proton) configuration for these nuclei would involve $d_{5/2}^3$. This leads to the unpaired configuration ($3/2^+$ state) becoming the ground state away from shell closures compared to the paired configuration ($5/2^+$ state) being the ground state for nuclei with a magic number of neutrons (protons). The switch of the ground state from $5/2^+$ to $3/2^+$ is due to the increasing importance of the long-range relative to pairing strength. From these, the $N=Z=$ 8, 14, and 16 shell closures are seen to be well established while the $N=20$ magicity has been eroded with the island of inversion clearly favouring the unpaired $3/2^+$ ground-state configuration. 

The limited study on $^{17}$C, $^{18}$N, $^{19}$O mentioned earlier~\cite{Wie2008b} suggests $^{17}$C to be an open nucleus.
The odd-odd $^{18}$N has a $1^-$ ground state with unpaired $d_{5/2}^3$ neutrons and is a transitional nucleus exhibiting increased importance of the long-range component as compared to $^{19}$O.

The location of the $\chi/G$ value for the $Z=4$ nucleus $^{15}$Be in Fig.~\ref{fig:systematic} (b) is indicative of significant uncertainties regarding the existence of the $3/2^+$ state as also noted in table~\ref{tab:protons}. Should the $5/2^+$ state indeed be the ground state in $^{15}$Be the model (which otherwise reproduces the global trends quite satisfactorily) would be contradicted since a spherical shell closure at $Z=4$ is difficult if not impossible to envisage. While it seems prudent to note that the spin and parity assignments of the ground and excited states in $^{15}$Be are tentative, we also realize that the resonance nature of these levels might limit the application of our model.  Nevertheless, further experimental investigations into the spin assignments of the states of  $^{15}$Be are desirable.

For the odd-even nuclei $N=Z=11$ the energy differences $E(5/2^+_1)-E(3/2^+_1)$ and $E(9/2^+_1)-E(3/2^+_1)$ clearly track the shell closures accurately, as shown in Fig.~\ref{fig:evenodd}, with $5/2^+$ ground states at closed-shells and $3/2^+$ ground states away from magicity as discussed above. As mentioned earlier, the notable exception is $^{33}$Na, which not only appears to exhibit a rotor behavior for $R_{39}$, but also an unexpectedly reduced $E(9/2_1^+)-E(3/2)_1^+$ value. This may be indicative that the tentative ($9/2^+$) spin assignment needs to be revisited, in particular, since the $R_{32}$ value, which is independent of $E(9/2_1^+)$ (data point at $\chi/G=10.15$ in Fig.~\ref{fig:model} (c)), does not exhibit the same behavior.
For $^{29}$Na the $E(5/2_1^+)-E(3/2_1^+)$ energy difference suggests that this nucleus with a $3/2^+$ ground state lies at the shores of the island of inversion, supporting the findings from $\beta$-decay studies \cite{Tri2005,Tri2006}.

\begin{figure}
\includegraphics[clip,angle=0,width=\columnwidth]{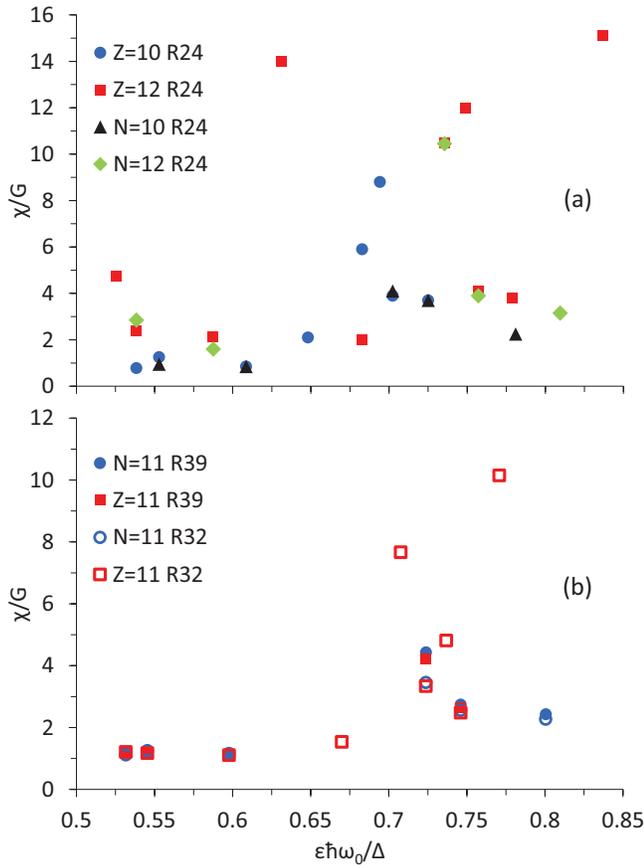}
\vspace{0cm}
\caption{(Color online) Correlation between $\chi/G$ and $\epsilon\hbar\omega_0/\Delta$ for (a) $R_{24}$ values from $d_{5/2}^{2,4}$ protons (blue solid circles and red solid squares) and neutrons (black solid triangles and green solid diamonds) and for (b) $R_{39}$ and $R_{32}$ values from a combination of $d_{5/2}^{2,4}$ and $d_{5/2}^{3}$ protons (red open and solid squares) and neutrons (blue open and solid circles). The data points for $^{15}$Be and $^{33}$Na have been omitted from panel (b). (see text for details)}
\label{fig:fig6}
\end{figure}

The overall systematic behavior is in agreement with the expectations that follow from the factor $P=N_p N_n/(N_p+N_n)$~\cite{Casten96} 
, which in heavier nuclei has been shown 
to be $\approx  2.8 \epsilon\hbar\omega_0/\Delta$. The dimensionless ratio $\epsilon\hbar\omega_0/\Delta$, the Migdal parameter, measures the competition between the deformation, $\epsilon$, and the pairing gap, $\Delta$, and is key in our understanding of the moments of inertia of nuclei~\cite{Migdal,BM}. In our approach, the ratio $\chi/G$ plays a similar role to $\epsilon\hbar\omega_0/\Delta$ and the two should be correlated.  This is actually seen in Fig.~\ref{fig:fig6} where we present the values of $\chi/G$ vs. the Migdal parameter, 
providing further support as to the applicability (at least qualitatively) of the effective interaction approach.
Note that $^{28}$Ne with $\epsilon\hbar\omega_0/\Delta = 0.65$ and $^{38}$Mg with $\epsilon\hbar\omega_0/\Delta = 0.63$ (panel (a)), and  $^{29}$Na with $\epsilon\hbar\omega_0/\Delta = 0.67$ (panel (b)) are located at the interfaces of the different structural regimes and exhibit  a transitional behaviour.


\section{Summary}
With the major advances we have witnessed in the development of large shell model and {\sl ab-initio} methods for nuclear structure, it is perhaps opportune to ask ourselves  about the value of the work presented above. Talmi's vision of explaining complex nuclei with simple models is still very much relevant as it usually offers an intuitive approach to understand and correlate structural information in terms of basic physical ingredients
(which sometimes get lost in the more sophisticated approaches).  Our analysis fits well within this context.

We have shown that a simple model for $d_{5/2}^{2,3,4}$ protons and neutrons,
with an effective-interaction that includes both short- and long-range terms is able to explain the experimental data with the appearance and disappearance of shell closures at $N=Z=8,14,16,20$. Furthermore, the trend in the competition between the quadrupole and pairing strengths, $\chi/G$, is in line with the $N_pN_n$-scheme expectations and appears  to be responsible for the inversion of the ground states in the odd-A systems.

\section{Acknowledgments}

This work is based on the research supported in part by the National Research Foundation of South Africa (Grant Number 118846) and by the Director, Office of Science, Office of Nuclear Physics, of the U.S. Department of Energy under Contract No. DE-AC02-05CH11231 (LBNL).

\end{document}